\RequirePackage[l2tabu, orthodox]{nag}
\documentclass{stsci_report}
\usepackage[pdftex,
pdfauthor={A. M. Guzman},
pdftitle={Analysis of the ACS/SBC from 2015-2024},
pdfsubject={},
pdfkeywords={ACS, CCD, Hubble Space Telescope, HST, Space Telescope Science Institute, STScI, Advanced Camera for Surveys, Dark Rate, Solar Blind Channel, SBC, SAA Passage, LODARK region}]{hyperref}
\usepackage[section]{placeins}
\usepackage{booktabs}
\usepackage{rotating}
\usepackage{natbib}
\usepackage{url}
\usepackage{graphicx}
\usepackage[font=footnotesize,labelfont=bf]{caption}
\usepackage{subcaption}
\usepackage{nameref}
\usepackage{hyperref} 
\usepackage{array} 
\usepackage{color}
\usepackage{amsmath}
\usepackage{amssymb}
\usepackage{gensymb}
\usepackage[bottom,symbol]{footmisc}

\usepackage{float}
\usepackage{aliascnt}
\newaliascnt{eqfloat}{equation}
\newfloat{eqfloat}{h}{eqflts}
\floatname{eqfloat}{Equation}
\usepackage{adjustbox}
\usepackage{threeparttable}
\usepackage{booktabs}

\copyrighttext{Copyright \copyright \the\year\ The Association of Universities for Research in Astronomy, Inc. All Rights Reserved.}

\title{\textbf{Updates to the SBC Dark Rate Monitor}}
\presubtitle{\flushleft{\includegraphics[width=5cm]{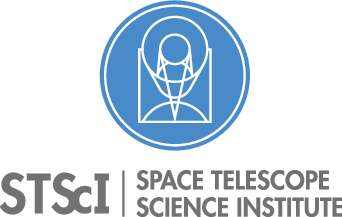}} \\ \hfill Instrument Science Report ACS 2024-04}
\author{A. M. Guzman, R. J. Avila}
\date{September 17, 2024}

\begin{document}

\maketitle

\abstract{The Solar Blind Channel (SBC) typically exhibits elevated dark rate levels at temperatures exceeding 25.5°C. However, instances of rapid dark rate increases have been observed before the detector reaches this threshold. To more closely monitor these anomalies, the existing calibration program was expanded, scheduling 24 total orbits per year distributed across eight visits for Cycle 31 and beyond. With enhanced data availability, we provide further updates and analysis of the dark rate in this report. We investigated the influence of the South Atlantic Anomaly (SAA) passage on dark rates. Using orbital parameters from SPT products, we plotted the Hubble Space Telescope's (HST) locations during dark exposures relative to the SAA. Comparison of HST paths during elevated and stable dark rate visits revealed no significant correlation between proximity to the SAA boundary and dark rate increases. Furthermore, we examined the stability of dark rates in the vicinity of the SBC-LODARK aperture within the detector. Our findings confirm that this region maintains consistently low dark rates across visits, unaffected by the elevated dark current observed elsewhere in the detector. The SBC-LODARK aperture therefore continues to be recommended for small sources.}

\section*{Introduction} \label{s:intro}
The Solar Blind Channel (SBC) in the Hubble Space Telescope's Advanced Camera for Surveys (HST/ACS) is a Multi-Anode Microchannel Array (MAMA) detector that exhibits low dark current at low temperatures. The SBC experiences elevated dark current in a region (approximately 750 x 750 pixels) located between the center and top right of the detector \citep{rob2018_isr}. There is a low dark rate region (approximately 390 x 390 pixels) at the lower left corner of the detector which is highly recommended for scientists to use to prevent increased dark noise from being present in their data. The dark rate in the SBC depends on the instrument's temperature and it has been shown that for this detector, the dark rate remains at approximately $ 8.52$ x $10^{-6}$ counts/s/pixel before reaching 25.5\degree C. It takes approximately two hours after the detector has been turned on for it to reach this temperature \citep{rob2017_isr}. The spatial distribution of the individual dark counts is random and changes from frame to frame, which is why there is no dark frame subtraction for MAMAs in the calibration pipeline CALACS.

Since 2002, the ACS Team has gathered dark images to analyze the dark rate in the detector and how it changes over time. Prior to Cycle 28 the dark rate in the SBC was only measured once a year and had remained low throughout this time frame. The dark images taken in Cycle 26 (2019-04-24) and Cycle 27 (2020-03-03) had unusually elevated dark rates which raised concerns over the status of the instrument. To investigate whether the SBC continued to exhibit these elevated dark rates the monitor program was expanded starting with Cycle 30. For Cycle 31 and beyond, we use 24 total orbits, eight visits per year, that consist of two types of visits. Long visits consist of twenty 1000-second exposures taken continuously over four orbits. Short visits consist of eleven 1000-second exposures taken continuously over two orbits. The long visits occur in December, March, June and September, while the short visits are taken in October, January, April and July, each visit being six weeks apart. We request that the detector be turned off for at least 10 hours before each visit is executed, however there have been visits where this condition was not met.

\section*{Data} \label{s:observations}
For this analysis, we obtained the RAW and SPT files from the Mikulski Archive for Space Telescopes (MAST). We calibrated all the RAW images using CALACS which returned FLT products. For RAW SBC dark data, CALACS converts the integer pixel values to floating point pixel values and writes temperature information to the FLT header. There is a numerical difference $<$ 0.01\% between non-zero values found in FLT and those in RAW files, introduced by converting from INT to FLOAT. There are no temperature sensors on the detector so we use the temperature measured in the enclosure tube as a proxy. The temperature is measured at the beginning and end of an exposure, and recorded in the MDECODT1 and MDECODT2 header keywords. We then measured the dark rate of each FLT file by calculating the mean value of the science frame (excluding pixels flagged as bad in the DQ array) and dividing it by the EXPTIME. Our calculations were then added to our data table. We used this data table to create the diagnostic and dark rate trend plots where we analyze whether the most recent visit is stable or elevated. An elevated visit is defined to be any visit where the dark rate began increasing prior to the instrument reaching 25.5\degree C and continued to increase exponentially. 

\begin{figure}[H]
  	\centering
 	\begin{subfigure}{1\textwidth}
  		\centering
		\includegraphics[scale=0.15]{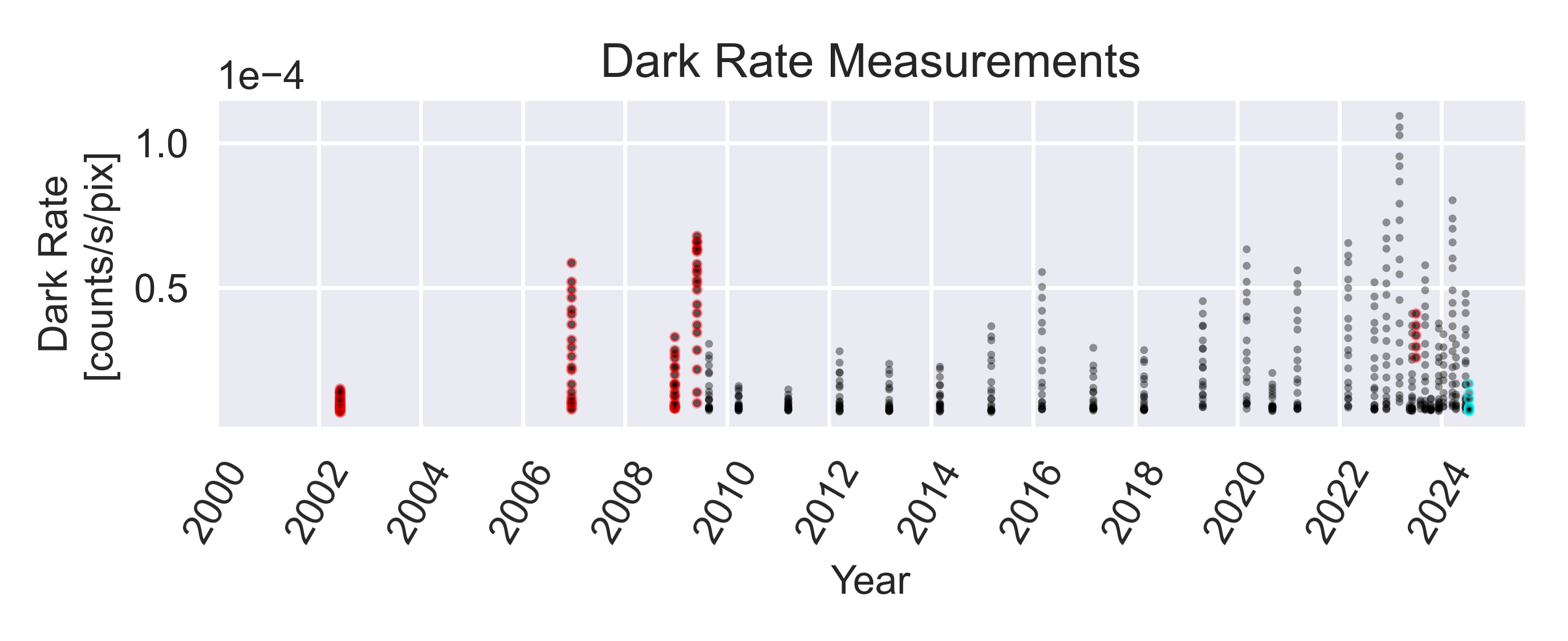}
		\end{subfigure}
 \caption{All the dark rate measurements over time. Each data point corresponds to the dark rate of a dark image during a visit. The red highlighted data represents the discarded visits not used for our analysis. The cyan highlighted data represents the most recent visit.}
    \label{fig:alldarkratetrend}
\end{figure}

There are a number of visits that are not used for our analysis, as shown in red on Figure \ref{fig:alldarkratetrend} and noted in Table \ref{tab:example}. The visit in Cycle 10 (PID: 9022, 2002-05-31) is not used for our analysis because it was an abnormally long visit taken during instrument checkout after launch. Both visits from Cycle 15 (PID: 11049, 2006-12-12) (PID: 11885, 2008-12-22) are discarded because HST passed through the SAA passage and therefore the dark rates were higher than usual \citep{rob2017_isr}. Hence, the Cycle 17 (PID: 11373, 2009-05-31) visit, and the 2023-07-01 visit from Cycle 30 Supplemental (PID: 17306) were also discarded because the detector had already been turned on, before the visit began, and the starting temperature was above the 25.5\degree C threshold. Table \ref{tab:all} contains the information for all visits to date.

\begin{table}[ht]
    \centering
    \begin{tabular}{|c|c|c|}
        \hline
        \textbf{Proposal ID} & \textbf{Observation Date} & \textbf{Notes} \\
        \hline
        9022 & 2002-05-31 & Abnormally long exposure time. \\
        \hline
        11049 & 2006-12-12 & HST passed through SAA passage. \\
        \hline
        11885 & 2008-12-22 & HST passed through SAA passage. \\
        \hline
        11373 & 2009-05-31 & Starting temperature was above 25.5\degree C. \\
        \hline
        17306 & 2023-07-01 & Starting temperature was above 25.5\degree C. \\
        \hline
    \end{tabular}
    \caption{Discarded visits due to different anomalies.}
    \label{tab:example}
\end{table}

\section*{Updates on Dark Rate} \label{s:updates}

\begin{figure}[H]
  	\centering
 	\begin{subfigure}{1\textwidth}
  		\centering
		\includegraphics[scale=0.15]{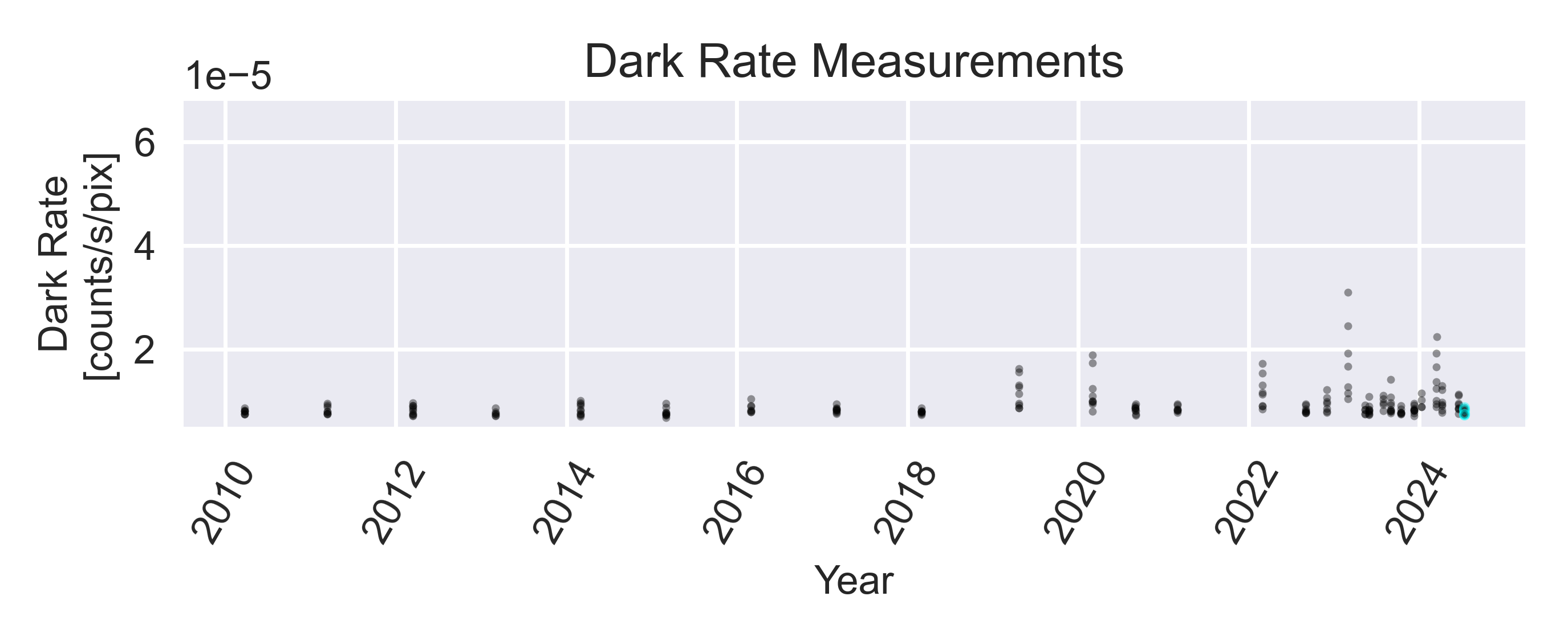}
		\end{subfigure}
 \caption{The dark rate measurements over time, when the temperature of the detector was below 25.5\degree C. Each data point corresponds to the dark rate of a dark image during a visit. The cyan highlighted data represents the most recent visit.}
    \label{fig:darkratetrend}
\end{figure}

We confined our data to only analyze dark images where the instrument's temperature was below 25.5\degree C, as seen on Figure \ref{fig:darkratetrend}. By having eight total visits per year we can investigate the frequency of high dark rate events. Since 2018, we have seen five elevated dark rate events, where the dark rate began increasing prior to the instrument reaching 25.5\degree C and continued to increase.

\begin{figure}[H]
  	\centering
 	\begin{subfigure}{1\textwidth}
  		\centering
		\includegraphics[scale=0.15]{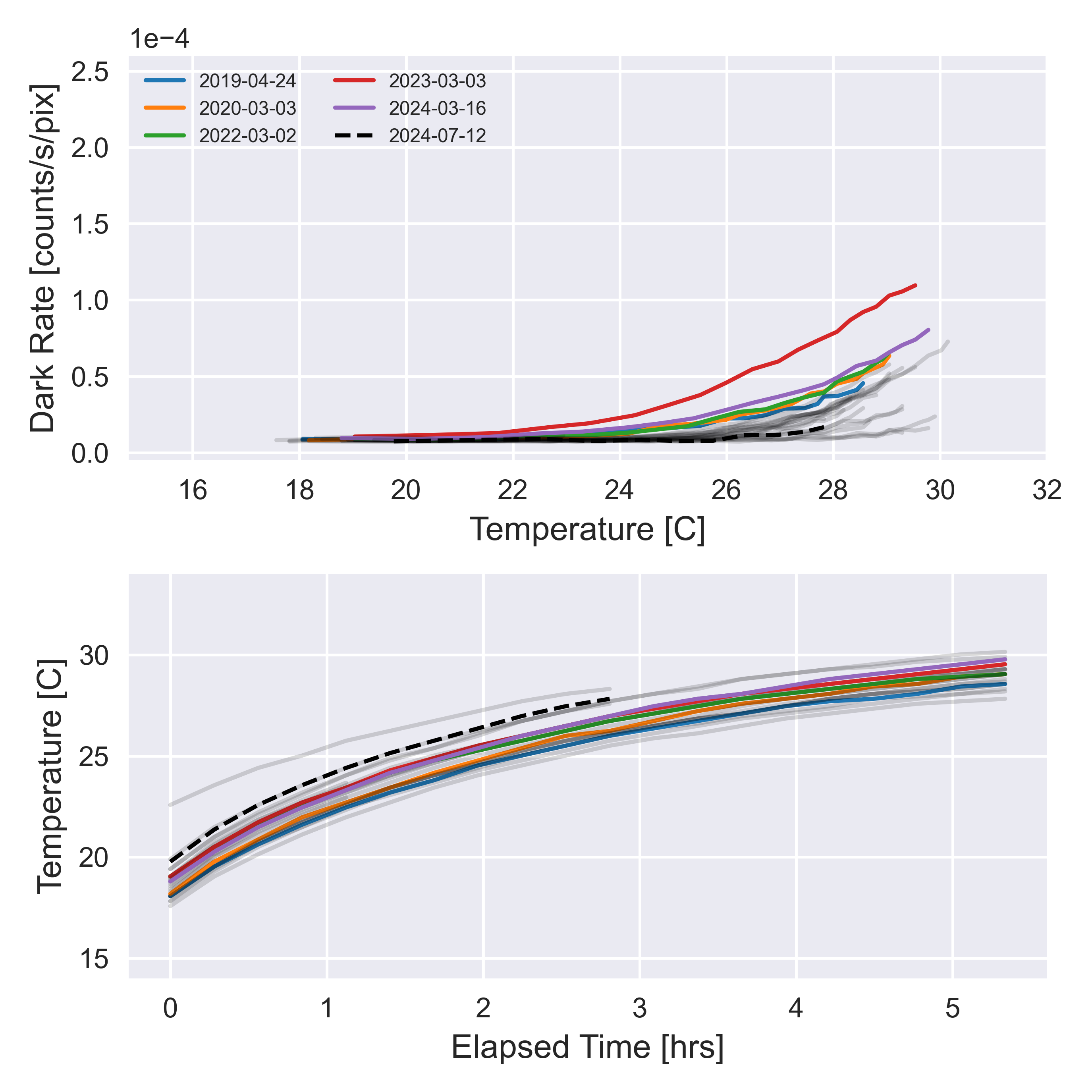}
		\end{subfigure}
 \caption{The top panel shows the relationship between the dark rate and temperature and the bottom panel shows the relationship between the temperature and elapsed time from when the detector was turned on. The dashed lines represent the latest visit and the most recent elevated visits are shown in different colors for both panels. The stable visits are plotted in grey.}
    \label{fig:diagnostic}
\end{figure}

These elevated visits are plotted in different colors in Figure \ref{fig:diagnostic}, where the top graph shows the dark rate vs temperature for all the datasets and the bottom graph shows the temperature vs elapsed time. The dashed lines in these plots represent the latest visit to date and the grey lines are the stable visits. All of the stable datasets plotted show how the dark rate remains very low until the instrument's temperature reaches 25.5\degree C whence the dark rate begins increasing rapidly. Looking at the bottom panel of Figure \ref{fig:diagnostic}, this temperature is reached approximately two hours after the SBC has been turned on. There are two datasets, the most recent 2024-07-13 visit and a stable visit, where the temperature in the detector was above its usual starting temperature. This only happens when the SBC detector was turned on for some time period prior to the dark visits. 

\begin{figure}[H]
  	\centering
 	\begin{subfigure}{1\textwidth}
  		\centering
		\includegraphics[scale=0.15]{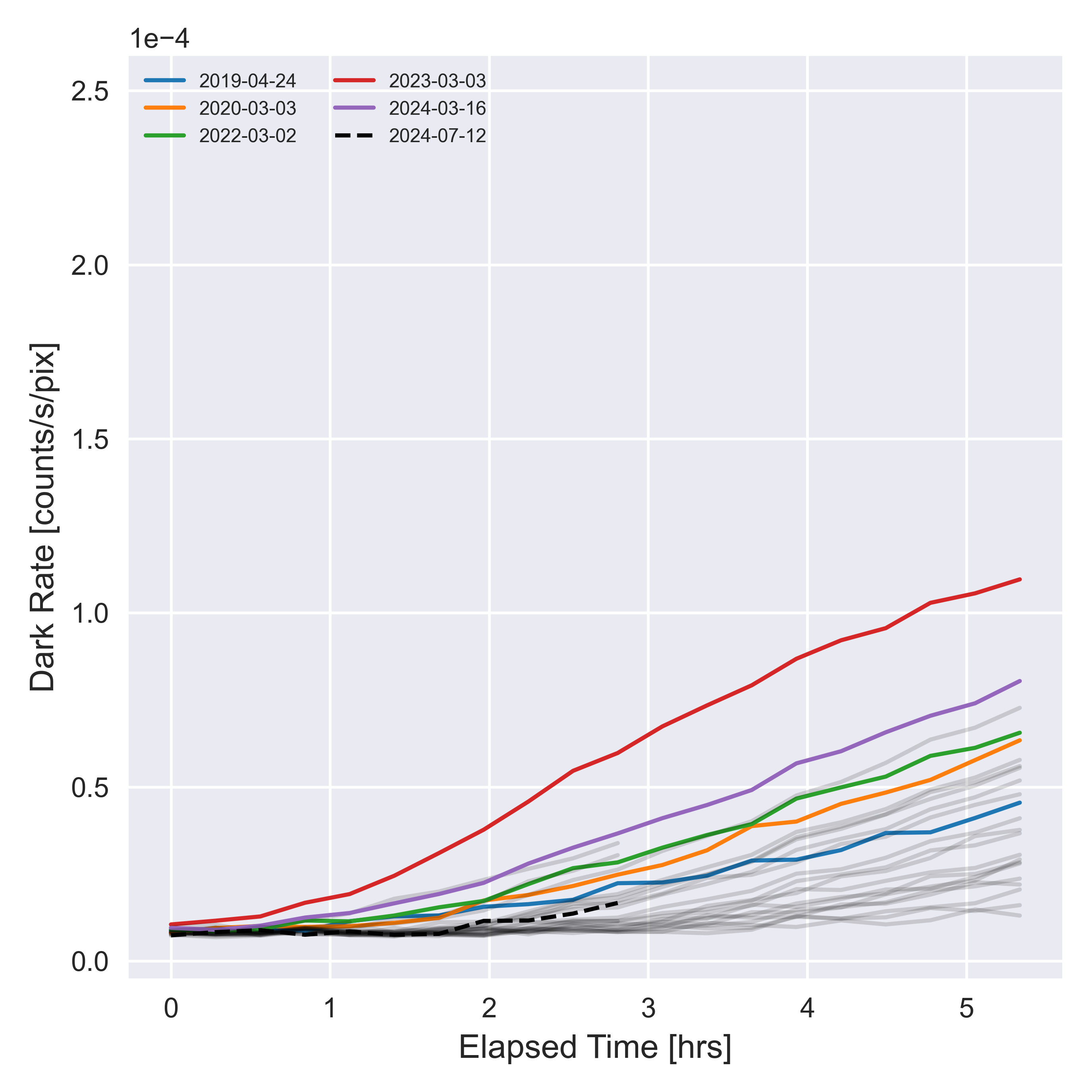}
		\end{subfigure}
 \caption{This plot shows the relationship between dark rate and the elapsed time after the detector has been turned on. The dashed lines represent the latest visit and the most recent elevated visits are shown in different colors for both panels. The stable visits are all plotted in grey.}
    \label{fig:darkelapsed}
\end{figure}

Figure \ref{fig:darkelapsed} gives us further insight in the behavior of the dark rate. The elevated visits, plotted in different colors, begin increasing either before or approximately at one hour after the detector has been turned on. No dependence is observed between dark rate and turn-on time.

\section*{South Atlantic Anomaly Passage} \label{s:saa}

The South Atlantic Anomaly (SAA) is a region where Earth's inner Van Allen radiation belt comes closest to the Earth's surface, causing increased particle flux, impacting satellite operations, like the SBC aboard HST, during close proximity to this area. The SBC does not operate while passing through the SAA, however, the HST does come very close to the SAA boundaries during some SBC visits. It was first suggested that the SBC experienced elevated visits when the HST crosses slightly over the SAA passage \citep{rob2017_isr}. We plotted the location of HST during both elevated and stable visits and compared their location relative to the SAA passage to see if the closer the HST is to the SAA passage the faster the dark rate increases. 

\begin{figure}[H]
  	\centering
 	\begin{subfigure}{1\textwidth}
  		\centering
		\includegraphics[scale=.4]{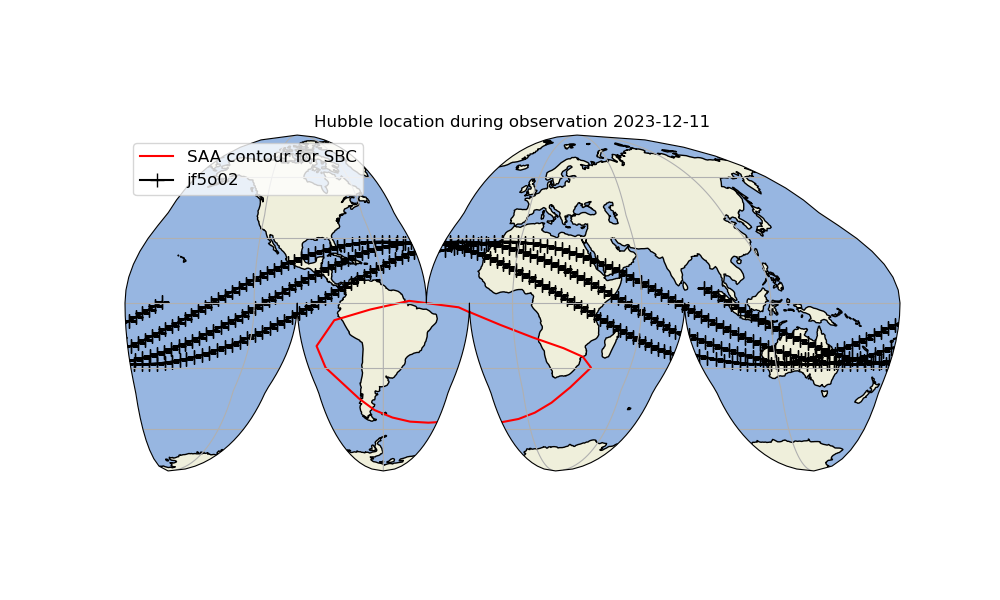}
		\end{subfigure}
 \caption{The location of HST during the non-elevated 2023-12-11 visit.}
    \label{fig:saa}
\end{figure}

\begin{figure}[H]
  	\centering
 	\begin{subfigure}{1\textwidth}
  		\centering
		\includegraphics[scale=.4]{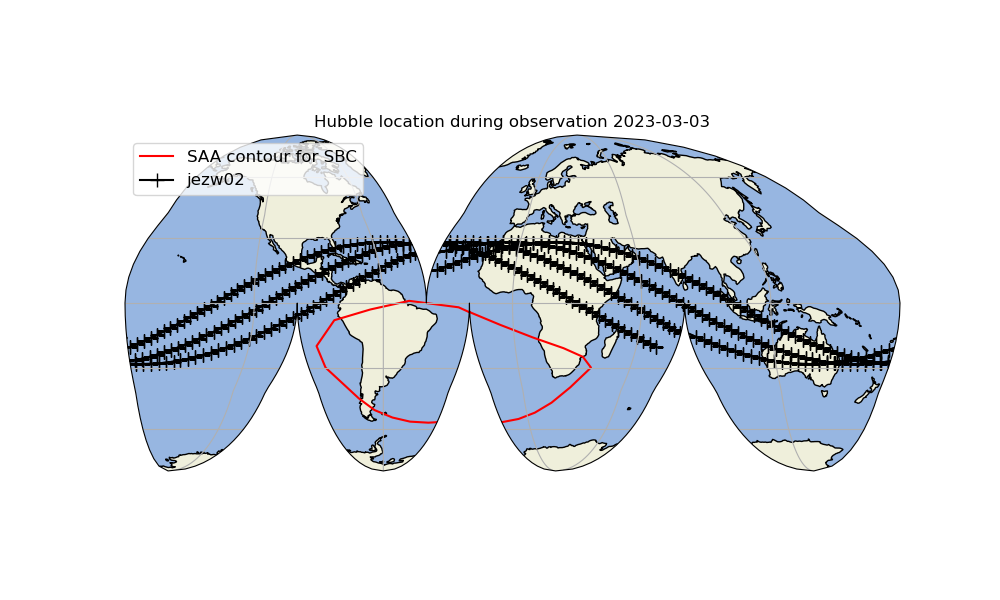}
		\end{subfigure}
 \caption{The location of HST during the elevated 2023-03-03 visit.}
    \label{fig:saahot}
\end{figure}

After analyzing both, a stable visit, Figure \ref{fig:saa}, and an elevated visit, Figure \ref{fig:saahot}, we came to the conclusion that SAA proximity is not the main cause for elevated dark rates seen in the SBC. Comparing both images, the distance the HST was from the SAA is fairly the same for both visits. We also discussed that, if the HST passing near the SAA caused the dark rate to increase exponentially at temperatures below 25.5\degree C, then we expect the dark rate to decrease slightly as the detector moves away from the SAA passage, which isn't the case as seen on Figure \ref{fig:diagnostic}. Although it is not shown in this report, we checked all the visits and found no correlation.

\section*{Low Dark and High Dark Regions} \label{s:lodarkregion}
\begin{figure}[H]
  	\centering
 	\begin{subfigure}{1\textwidth}
  		\centering
		\includegraphics[scale=0.24]{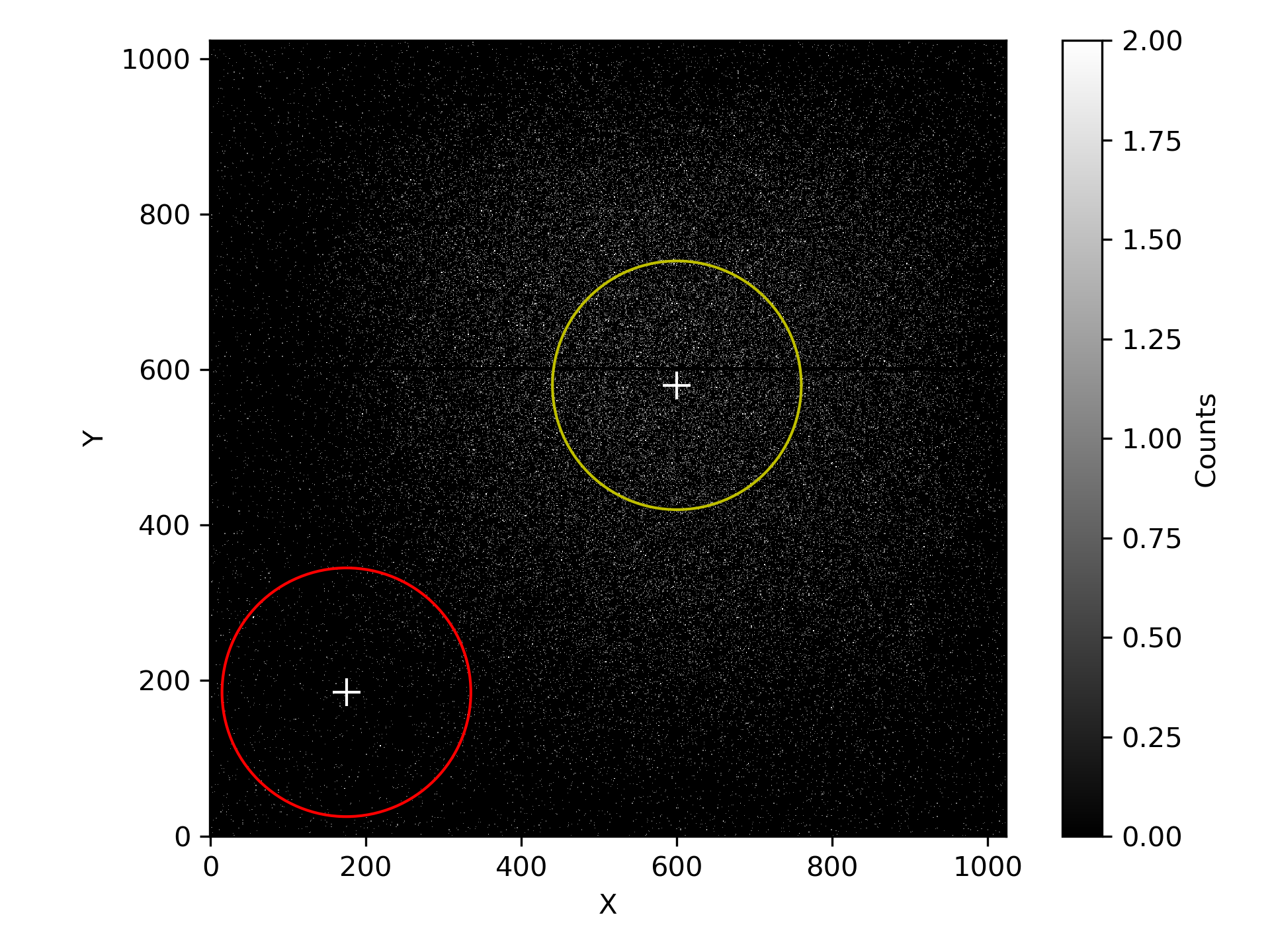}
		\end{subfigure}
 \caption{A dark frame exposure from the 2020-03-03 elevated visit (je5x01kbq FLT file). The low dark rate region is circled in red, with the center being at 175, 185 and a radius of 160 pixels. The high dark rate region is circled in yellow, with the center being at 600, 580 and a radius of 160 pixels.}
    \label{fig:lo_region}
\end{figure}

The dark rate in the SBC is elevated in the middle and upper-right corner of the detector, however the lower-left corner, circled red in Figure \ref{fig:lo_region}, does not exhibit these elevated dark rates. This SBC-LODARK aperture was enabled in order to provide observers with a convenient location to use if they required low dark rates \citep{rob2017_isr}. With the added data gained from the expanded calibration program we can confirm that this region continues to be the ideal area for observing as the dark rate remains low. We also analyzed the high dark current region, circled yellow in Figure \ref{fig:lo_region}, to investigate how high the dark rate gets in this region within a visit. 

\subsection*{Low Dark Region}

\begin{figure}[H]
  	\centering
 	\begin{subfigure}{1\textwidth}
  		\centering
		\includegraphics[scale=0.15]{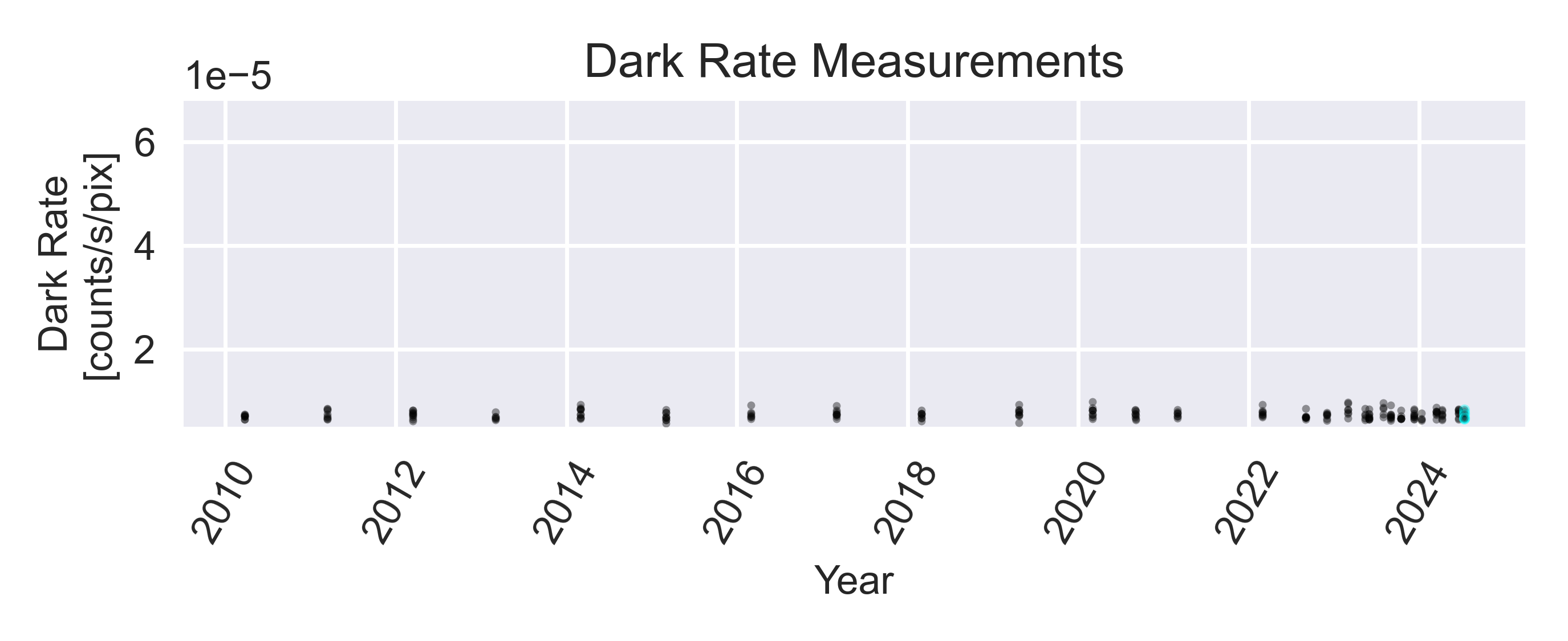}
		\end{subfigure}
 \caption{Same as Figure \ref{fig:darkratetrend} but for the low dark rate region.}
    \label{fig:lo_darkratetrend}
\end{figure}

Figure \ref{fig:lo_darkratetrend} illustrates the consistent stability of dark rates in the vicinity of the SBC-LODARK aperture since 2002. Even during visits when the overall detector's dark rate is elevated, the low dark rate region consistently maintains at low levels of dark current. 

\begin{figure}[H]
  	\centering
 	\begin{subfigure}{1\textwidth}
  		\centering
		\includegraphics[scale=0.14]{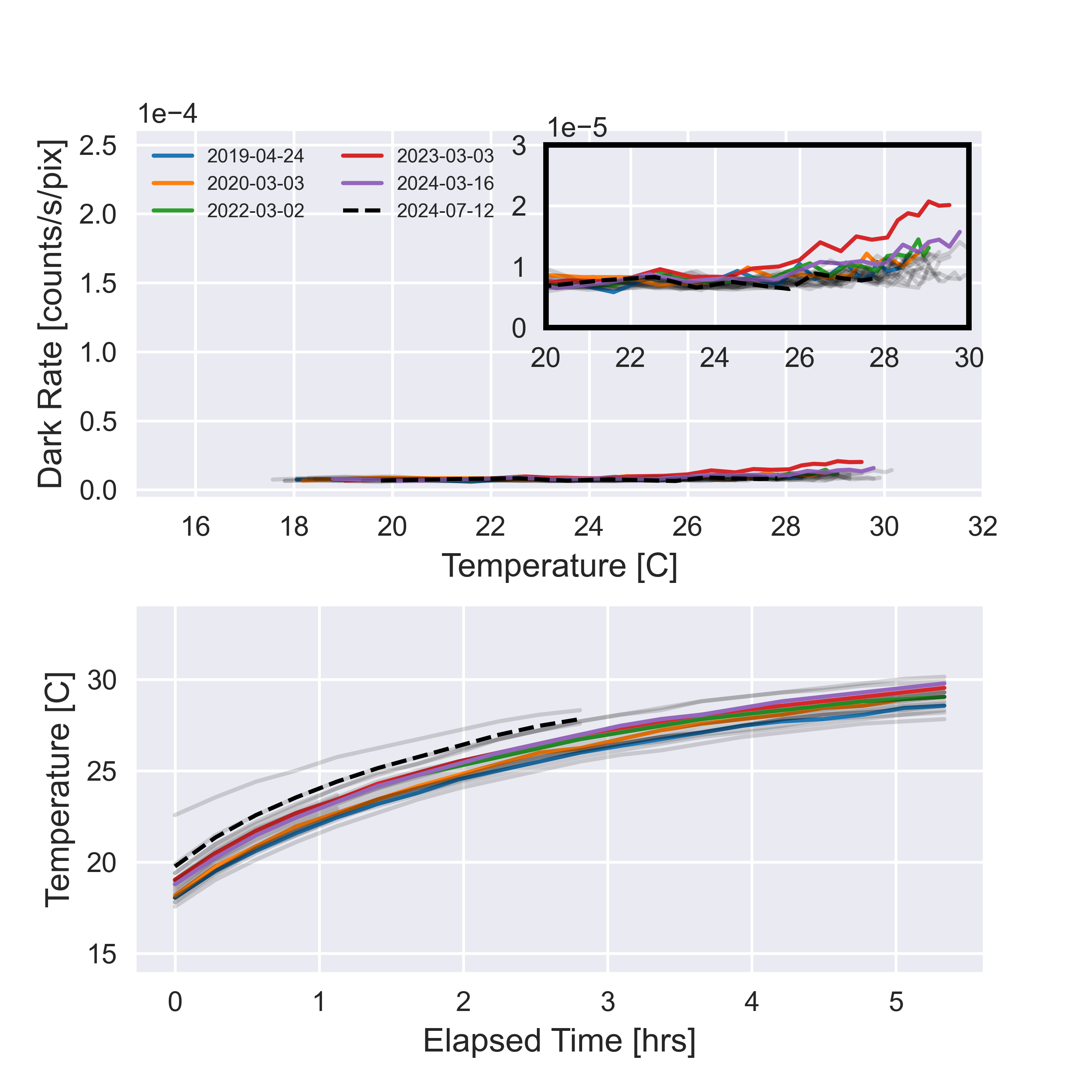}
		\end{subfigure}
 \caption{Same as Figure \ref{fig:diagnostic} but for the low dark rate region. The box is a zoomed in version of the data for clarity.}
    \label{fig:lo_diagnostic}
\end{figure}

\begin{figure}[H]
  	\centering
 	\begin{subfigure}{1\textwidth}
  		\centering
		\includegraphics[scale=0.14]{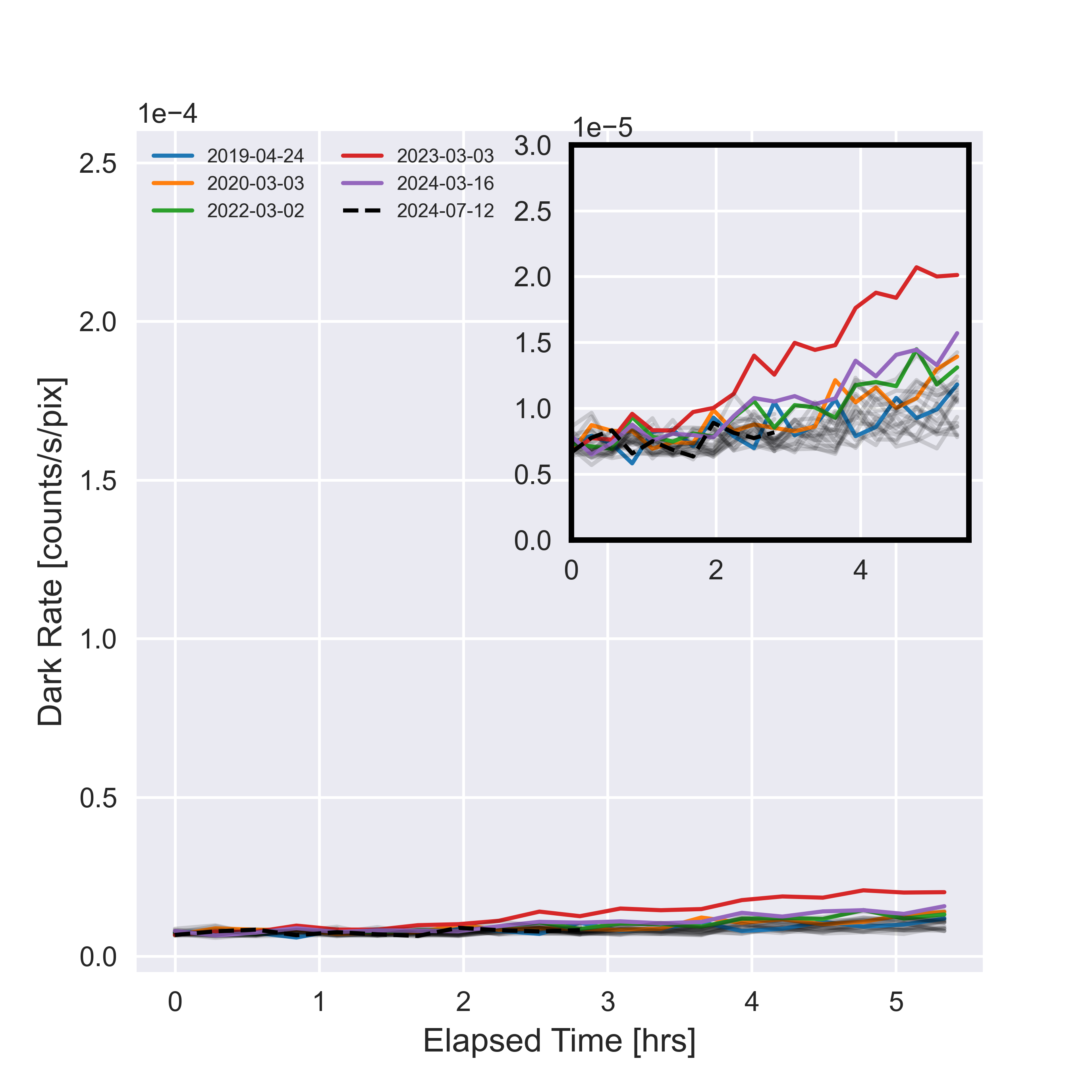}
		\end{subfigure}
 \caption{Same as Figure \ref{fig:darkelapsed} but for the low dark rate region. The box is a zoomed in version of the data for clarity.}
    \label{fig:lo_darkelapsed}
\end{figure}

In Figure \ref{fig:lo_diagnostic}, the top panel displays the plot of dark rate against temperature, while the bottom panel illustrates the relationship between temperature and elapsed time. Notably, the visit on 2023-03-03 stands out as the only instance where the dark rate in the low dark region exceeded the norm. However, even during this visit, the dark rate remained an order of magnitude lower than the overall detector's dark rate. Subsequent elevated visits depicted in the plot appear to exhibit stability, aligning with similar patterns observed during stable visits. These trends are also seen in the dark rate versus elapsed time in Figure \ref{fig:lo_darkelapsed} which further supports our findings.

\subsection*{High Dark Region}
The high dark region we chose to analyze is a circular region centered at 600, 580 that experiences elevated dark current compared to the overall detector. We performed an analysis of this region to provide users with further information of the dark rate trends in this region.

\begin{figure}[H]
  	\centering
 	\begin{subfigure}{1\textwidth}
  		\centering
		\includegraphics[scale=0.15]{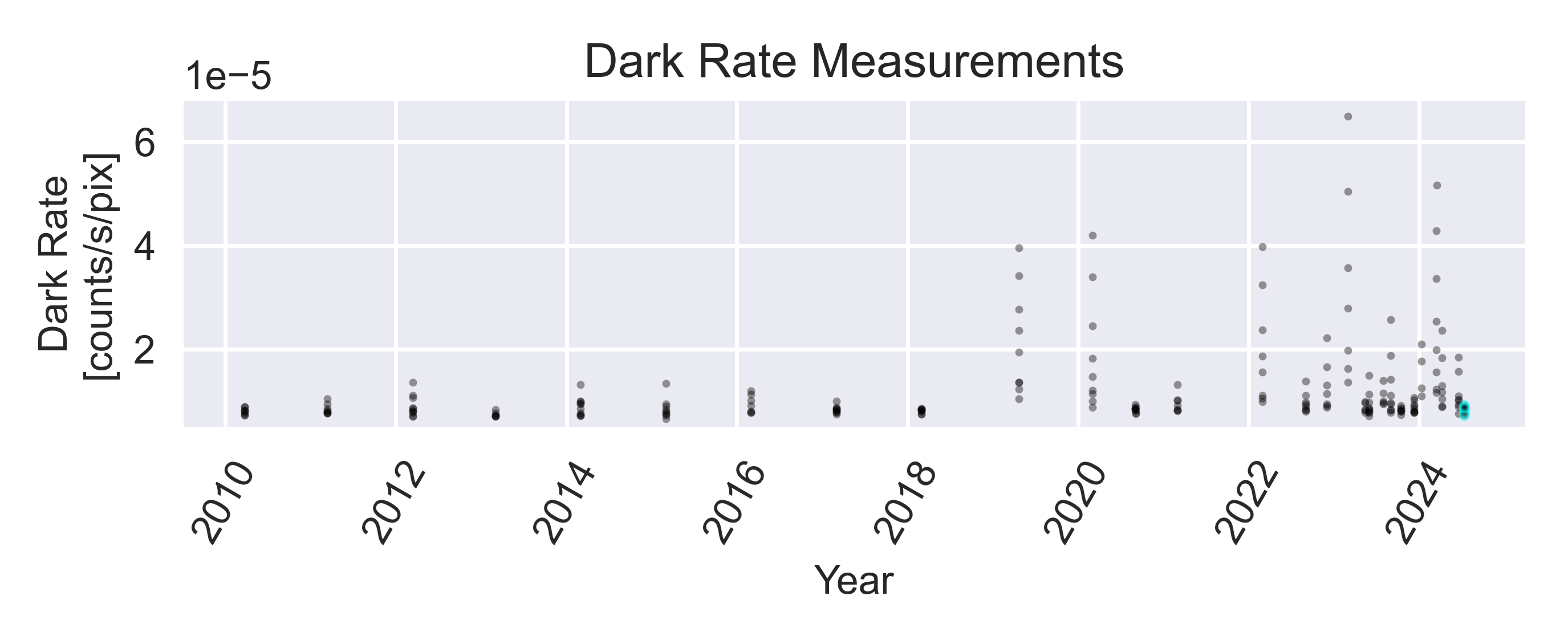}
		\end{subfigure}
 \caption{Same as Figure \ref{fig:darkratetrend} but for the high dark rate region.}
    \label{fig:high_darkratetrend}
\end{figure}

\begin{figure}[H]
  	\centering
 	\begin{subfigure}{1\textwidth}
  		\centering
		\includegraphics[scale=0.15]{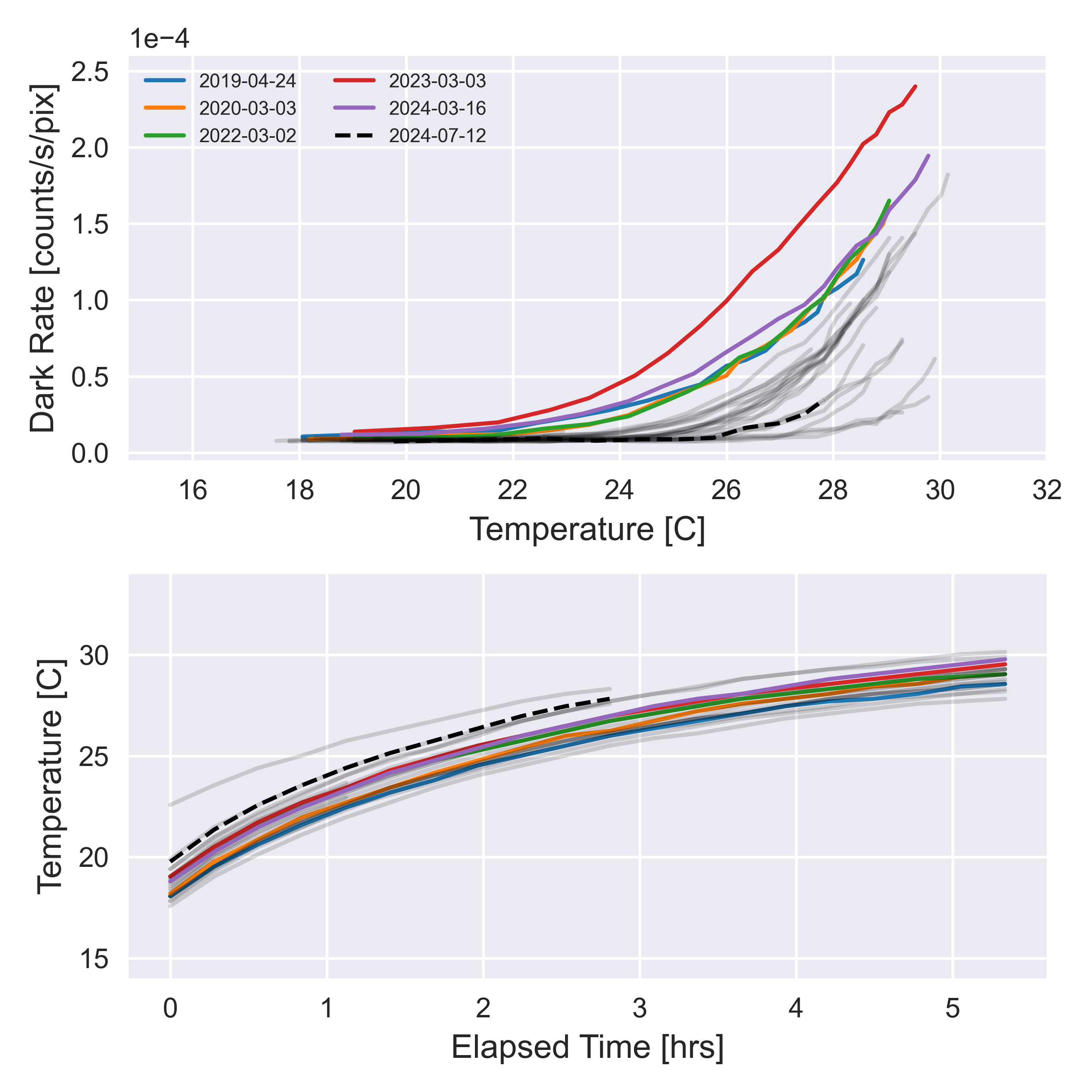}
		\end{subfigure}
 \caption{Same as Figure \ref{fig:diagnostic} but for the high dark rate region.}
    \label{fig:high_diagnostic}
\end{figure}

\begin{figure}[H]
  	\centering
 	\begin{subfigure}{1\textwidth}
  		\centering
		\includegraphics[scale=0.14]{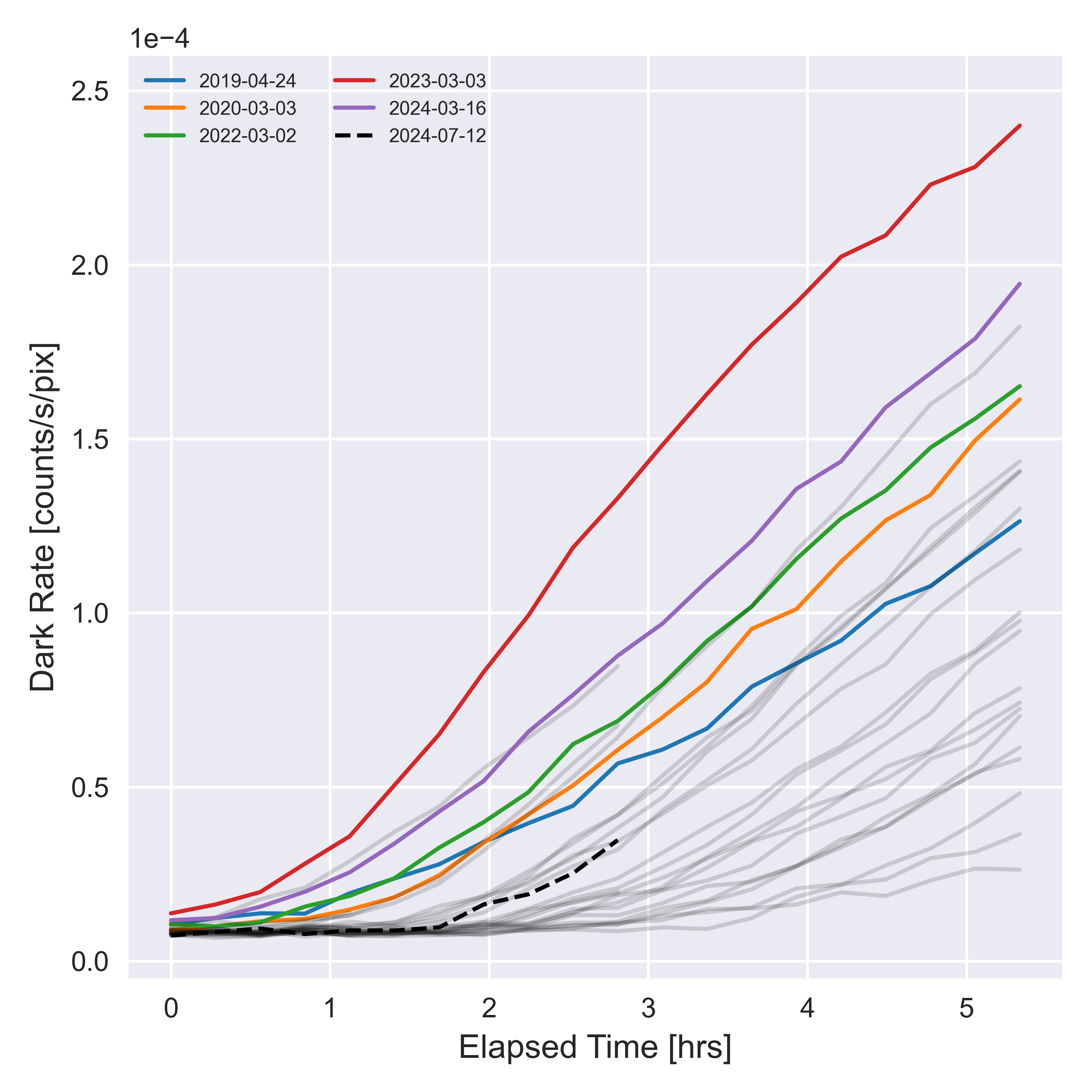}
		\end{subfigure}
 \caption{Same as Figure \ref{fig:darkelapsed} but for the high dark rate region.}
    \label{fig:high_darkelapsed}
\end{figure}

In the first graph of Figure \ref{fig:high_darkratetrend}, there is a clear separation between stable and elevated visits, with the dark rate increasing at approximately 22.5\degree C for elevated visits. Both Figures \ref{fig:high_darkratetrend} and \ref{fig:high_diagnostic} show the elevated dark rates in this high dark region. We decided to include Figure \ref{fig:high_darkelapsed} to show that in the high dark rate region, there's still no correlation between dark rate and elapsed time from when the detector was turned on. The dark rate measurements show how even during stable visits, this region's dark rate is excessively high and therefore contributes to noise in the image.

We recommend for users to position their target on the SBC-LODARK aperture whenever possible to avoid high dark current. 

\section*{Conclusion} \label{s:conclusion}

The SBC experiences high dark rate levels at temperatures above 25.5\degree C, however, there have been visits where the dark rate increased rapidly prior to the detector reaching this temperature. In response to several of these events, the calibration program was expanded to increase the number of visits per year to 24 total orbits, distributed across eight visits per year for Cycle 31. With more data we are now able to further investigate why certain visits experience higher dark rates at temperatures below 25.5\degree C. 

We investigated whether the SAA passage influenced the dark rate. For this we used the orbital parameters obtained from the SPT products, to plot the location of the observatory during dark exposures, in particular its location with respect to the SAA. We compared the HST path from elevated and stable visits to see whether the telescope's position in relation to the SAA boundary increased the dark rate. We conclude that proximity to the SAA boundary did not cause these elevated dark rates as the locations of the HST in both visits were similar.

We also explored whether the region of the detector near the LODARK aperture maintains a stable, low dark rate. We expect this region to have stable dark rates throughout every visit, and with the new data we were able to confirm that this region is not affected by the elevated dark current in the center of the detector. Therefore, it is advised for observers to use this region for their proposed target whenever possible.

\section*{Acknowledgements}

We thank Norman Grogin for assistance, feedback and support through-out the project. We also thank the following ACS team members for providing comments to improve this report: David Stark and Jenna Ryon. 

For this report we used the following: \texttt{jupyter} \citep{jupyter}, \texttt{numpy} and \texttt{scipy} \citep{cite_numpy}, \texttt{pandas} \citep{pandas}, \texttt{astropy} \citep{astropy}, and  \texttt{matplotlib} \citep{matplotlib}.

\bibliography{guzman_sbc}
\bibliographystyle{apj}

\section*{Appendix}

\begin{table}[ht]
    \centering
    \begin{tabular}{|c|c|c|}
        \hline
        \textbf{Proposal ID} & \textbf{Observation Dates} & \textbf{Notes} \\
        \hline
        9022 & 2002-05-31 & Abnormally long exposure time. \\
        \hline
        11049 & 2006-12-12 & HST passed through SAA passage. \\
        \hline
        11885 & 2008-12-22, 2009-08-13, & 2008-12-22: HST passed through \\
        		&	2010-03-22 & SAA passage. \\
        \hline
        11373 & 2009-05-31 & Starting temperature was  \\
         & & above 25.5\degree C. \\
        \hline
        12391 & 2011-03-13 &  \\
        \hline
        12736 & 2012-03-12 & \\
        \hline
        13161 & 2013-03-03 & \\
        \hline
	13598 & 2014-03-01 & \\
	\hline
	13961 & 2015-03-03 & \\
	\hline
	14404 & 2016-03-01 & \\
	\hline
	14513 & 2017-03-01 & \\
	\hline
	14955 & 2018-03-01 & \\
	\hline
	15528 & 2019-04-24 & Elevated visit at temperatures\\ 
	 & & below 25.5\degree C. \\ 
	\hline
	15766 & 2020-03-03 & Elevated visit at temperatures\\ 
	 & & below 25.5\degree C. \\ 
	\hline
	16386 & 2020-09-03, 2021-03-01 & \\
	\hline
	16529 & 2022-03-02 & Elevated visit at temperatures\\ 
	 & & below 25.5\degree C. \\ 
	\hline
	16977 & 2022-09-03 through 2023-09-01 & 2023-03-03: Elevated visit at temperatures\\ 
	 & & below 25.5\degree C. \\ 	
	 \hline
	17306 &  2023-05-13 through 2023-08-01 & 2023-07-01: Starting temperature was \\
	 & & above 25.5\degree C. \\
	\hline
	17340 & 2023-10-14 through 2024-09-01 & 2024-03-16: Elevated visit at temperatures\\ 
	 & & below 25.5\degree C. \\ 
	 \hline
    \end{tabular}
    \caption{All visits are included in the database for completeness.}
    \label{tab:all}
\end{table}

\end{document}